# Impact of the III-V/Ge nucleation routine on the performance of high efficiency multijunction solar cells

Laura Barrutia[1*], Iván García[1], Enrique Barrigón[2], Mario Ochoa[3], Carlos Algora[1]

and Ignacio Rey-Stolle[1]

[1] Instituto de Energía Solar, Universidad Politécnica de Madrid, Avda. Complutense 30, Madrid, Spain

[2] Solid State Physics and NanoLund, Lund University, P.O. Box 118, SE-221 00 Lund, Sweden

[3] Laboratory for Thin Films and Photovoltaics, Empa-Swiss Federal Laboratories for Materials Science and Technology, Ueberlandstrasse 129, 8600 Duebendorf, Switzerland

*email: laura.barrutia@ies-def.upm.es

**Abstract**

This paper addresses the influence of III-V nucleation routines on Ge substrates for the growth of high efficiency multijunction solar cells. Three exemplary nucleation routines with differences in thickness and temperature were evaluated. The resulting open circuit voltage of triple-junction solar cells with these designs is significantly affected (up to 50 mV for the best optimization routine), whereas minimal differences in short circuit current are observed. Electroluminescence measurements show that both the Ge bottom cell and the Ga(In)As middle cell present a $V_{OC}$ gain of 25 mV each. This result indicates that the first stages of the growth not only affect the Ge subcell itself but also to subsequent subcells. This study highlights the impact of the nucleation routine design in the performance of high efficiency multijunction solar cell based on Ge substrates.

**Introduction**

The majority of the solar cells used today in space and CPV applications are multijunction solar cells grown on germanium substrates [1]. Among these, the main designs include the lattice matched GaInP/Ga(In)As/Ge triple-junction solar cell (3JSC) [2], where top and middle cell are grown lattice-matched to the Ge substrate (i.e. In content in the Ga(In)As is ~1%), and the upright metamorphic GaInP/GaInAs/Ge 3JSC, where a certain lattice mismatch of ~1.2% is allowed in both GaInP and GaInAs subcells (i.e. with a higher indium content of ~17% in the GaInAs subcell) in order to reach a better combination of bandgaps [3, 4]. In these kinds of solar cells, Ge substrate is both the mechanical support of the whole stack and act as the Ge bottom cell (BC). Therefore, the characteristics of the Ge BC result from the heteroepitaxy processes used for its formation (i.e. the nucleation) and the indirect processes that the subcell needs to withstand during the growth of the rest of the structure (in particular, the thermal load associated with the growth of the upper subcells and tunnel junctions)[5].

To approach optimum performance in these multijunction solar cells (MJSC), it is necessary that each subcell realizes its maximum potential. In this context, it is known that the Ge BC is the weakest link in a MJSC. First, Ge has a bandgap significantly lower than the optimum for a 3JSC and therefore, it provides an excess in photocurrent ($J_{SC}$) with respect to the upper subcells, which cannot be used. The way to optimize the use of the solar spectrum is to upgrade to a 4J device with a 1eV junction between the Ge and Ga(In)As subcells [6, 7]. Second, Ge yields low $V_{OC}$ in comparison with similar band gap p-n junctions (i.e; InGaAs p-n junction of ~ 0.74eV). Ge subcells tend to produce low values with typical bandgap-voltage offsets ($W_{OC} = E_g/q - V_{OC}$) in the range of 0.40-0.45V [5], suggesting that there is margin for improvement.

The growth of a MJSC is a multiparametric process entailing several challenges that directly or indirectly affect Ge subcell performance. First, a heteroepitaxial routine is required to grow defect-free polar III-V layers on a non-polar Ge substrate without antiphase domains yielding a high quality nucleation layer (i.e. with good morphology and defect-free) for the subsequent growth of other layers of the whole MJSC structure. Second, the III-V in-diffusion and Ge out-diffusion into/from the substrate need to be minimized. The Ge BC is formed by the in-diffusion of group-V and group-III elements into the p-type Ge wafer during the growth of the nucleation layer, typically made of GaAs or GaInP. One of the main challenges when optimizing the Ge BC performance is to control the emitter depth which is the direct result of the diffusion profile of both group-III and V elements. Besides, there is also Ge out-diffusion into the III-V material that needs to be considered to prevent the compensation of active layers (for example in tunnel junctions). In addition, there is Ge gas-phase autodoping that can be incorporated into the III-V epilayers as an unintended background dopant. On top of all these challenges, once the Ge BC is formed, the epitaxial growth continues for 1 to 2 hours at 640°C-665°C (typically) to complete the upper subcells and tunnel junctions, which implies an extra thermal load that the Ge BC needs to deal with. In summary, the III-V MJSC growth determines and affects the Ge BC properties and performance.

Some of the aforementioned problems or trade-offs have been studied by different research groups over the last two decades [8-10]. For example, the use of misoriented substrates [11] in conjunction with specific surface preparation routines [12-15] have been widely employed to grow defect-free III-V nucleation layers on Ge substrates. Another example of surface preparation, in this case for the growth of GaAs on Ge by MBE for laser diode applications, was the use of a Ga pre-layer technique that favors smoother surface morphology [16]. Regarding the negative effect of the Ge solid-phase out-diffusion, diffusion barriers with AlAs or GaAs epilayers have been proposed [17,18]. The first method consists on the growth of a thin AlAs layer that acts as a barrier blocking this cross diffusion and this effect was attributed to a higher Al-As bonding energy [17]. The disadvantage of this process is the high oxygen incorporation in Al-containing alloys that could lead to highly resistive layers. Another solution developed in the past was the growth of a thin GaAs nucleation layer at low temperature that acts as a barrier for Ge [18]. Anyhow, we must point out that all these experiments were devised for GaAs nucleation layers grown on inactive Ge substrates. When it comes to the optimization of the Ge BC, GaInP nucleation layers are preferred since they have been shown to produce shallower emitters than their GaAs counterparts. This is so because phosphorus diffusion coefficient in Ge is lower than that of arsenic [19]. Therefore, we have used GaInP nucleation layers in this work, which is an advantage for the specific requirements of the emitter of the Ge BC. Additionally, it has been shown that other emitter parameters —such as doping, diffusion length and surface recombination velocity— also play an important role in the $V_{OC}$ of the cell [5, 8]. The variables affecting Ge autodoping and how to mitigate it have been also studied and evaluated in several works [20-22]. Finally, the extra thermal load that the Ge BC suffers has been demonstrated to affect significantly the $V_{OC}$ of the devices. In [5], it was shown that Ge cells grown as single junction devices had a $V_{OC}$ notably higher than that of the same design acting as a BC in a 3JSC ($V_{OC}$ dropped from ∼ 250 mV down to <200 mV).

Here, we want to highlight another aspect affecting Ge-based MJSC performance. In particular, the aim of this work is to study the impact of the nucleation layer growth parameters in 3JSCs performance. We have designed three nucleation routines, with considerable differences in thermal load (a conservative, a moderate and an aggressive design). These three designs have been applied to grow complete lattice-matched GaInP/Ga(In)As/Ge solar cells (3JSCs). In order to understand the influence of these nucleation designs, single Ge BCs devices were also analyzed. In this way, the sole impact of these designs in initial single junctions and subsequently in complete 3JSCs can be decoupled. This work might also help to understand the potential gains in $V_{OC}$ of MOVPE processes using high growth rates and therefore minimizing the thermal load [23, 24], as well as the possibility of using lower growth temperature ranges preserving the good quality of the epilayers.

**Experimental**

Lattice-matched GaInP/Ga(In)As/Ge solar cells were grown in a low pressure AIX 200/4 MOVPE reactor. $AsH_3$, $PH_3$, TMGa, TMAl, TMIn, DETe, DMZn, $DTBSi_2$ and $CBr_4$ were used as constituent and dopant precursors. P-type, gallium doped, 2-inch Ge (100) wafers with a 6º misorientation towards the [111] plane were used. Further details about the epitaxial growth can be found elsewhere [13].

Three different nucleation routines have been used for this study. Growth parameters such as growth rates and V/III ratios were kept the same for the three designs. GaInP layers were grown with a V/III ratio of ∼ 400 and a growth rate of ∼ 0.21 nm/s while GaInAs layers were grown with a V/III ratio of ∼ 10 and a growth rate of ∼ 0.83 nm/s. The differences in nucleation designs were mainly focused on both GaInP and Ga(In)As layer thickness together with Ga(In)As layer growth temperature. A conservative nucleation routine was designed with a GaInP layer of ∼350 nm and a thick Ga(In)As buffer of ∼1000 nm. The goal of such thick Ga(In)As layer is to preserve good morphology, guarantee a surface of the best quality, free of dislocations and to prevent the Ge solid-phase diffusion from reaching the upper active layers. Hereinafter, this semiconductor structure (depicted in Figure 1) will be called as *routine A*. The other two nucleation routines explored in this work look for the mitigation of the thermal load during the growth of these initial layers with shorter growth times and lower temperatures. Nucleation *routine B* basically consists in an intermediate design with a reduction of the thickness of both the GaInP nucleation layer and Ga(In)As buffer layer along with the decrease of the growth temperature for the Ga(In)As buffer. Finally, nucleation *routine C* presents a similar thickness reduction of the GaInP layer together with a more aggressive reduction of the thickness of the Ga(In)As buffer layer (from 1000 nm down to 50 nm). Table 1 summarizes the parameters used for the three designs: A (conservative), B (intermediate) and C (aggressive).

For nucleation routines B and C the thickness of the GaInP layer was set to a nominal value of ~ 50 nm (see Table 1). This value is high enough to avoid diffusion of arsenic coming from the Ga(In)As buffer layer into the Ge, which has been reported to occur with thicknesses below 35 nm [25]. If this were the case, the arsenic would partner with phosphorus in the formation of the BC emitter, which would result in deeper overall diffusion of group-V elements yielding thicker emitters and thus lower performance [8].

Details on the development of our complete 3JSCs by MOVPE can be found in [26]. The thermal load associated with the top and middle subcell growth was analogous for the three epitaxial structures (75 min for the middle cell growth at 640 °C + 42 min for the top cell growth at 665 °C).

Solar cell devices with an area of 0.11 cm$^2$ were used to determine the external quantum efficiency and dark J-V of as grown Ge solar cells (1JSCs). Samples were processed using standard photolithographic techniques. The front grid and back contact were formed using electroplated gold and devices were electrically isolated by wet chemical etching. The Ga(In)As buffer layer was chemically etched from the active area in 1JSCs using $NH_4OH:H_2O_2:H_2O$ (2:1:10). The GaInP layer was left to be used as the window layer of the BC. For the analysis of 3JSCs an analogous fabrication process was used with a solar cell device area of 0.06 cm$^2$. No antireflection coating was deposited on the samples.

Dark I-V curves were measured using the four-wire method with a Keithley 2602 source-meter instrument. The setup employed to measure the External Quantum Efficiency (EQE) is based on a 1000 W Xe lamp and a grating monochromator. Further details on this system and the measurement procedure can be found in [27]. A calibrated detector was used to measure the specular reflectance of the devices. The Internal Quantum Efficiency (IQE) was calculated from the EQE and the reflectance of the solar cells without anti-reflection coatings (ARC) following equation 1.

$$IQE = EQE / (1-R) \qquad \text{Eq 1}$$

The short-circuit current ($J_{SC}$) is calculated from the convolution of the IQE and the AM1.5d G173 standard spectrum. Measurements under concentrated light were performed with a flash simulator. For a detailed explanation of the set-up for concentration measurements, the flash lamp and its potential, the reader is referred to [28].

Secondary Ion Mass Spectroscopy (SIMS) with $Cs^+$ primary ions detection and negative ions detection was used for the quantification of Ge atoms in the first two semiconductor layers (GaInP nucleation and Ga(In)As buffer layer).

Electroluminescence (EL) spectroscopy was used as a no-destructive technique to obtain individual access to the electrical performance of each subcell in the triple junction solar cells. EL measurements together with EQE measurements were used to reproduce the individual dark I- V curve of each subcell. EL measurements were taken at NREL using a calibrated fiber-based spectrometer to sample the emitted light (the details can be found elsewhere [29]).

**Results and Discussion**

As a preamble, the impact of III-V nucleation routines was initially studied on single Ge devices (1JCs). In this way, the pure influence of these designs on Ge BCs prior to the thermal load associated with the growth of upper subcells can be assessed. The electrical performance was studied by means of Internal Quantum Efficiency (IQE) and dark J-V curves.

Figure 2 shows the IQE for nucleation A, B and C. The changes observed from 350-685 nm approximately are related to different optical thickness, i.e., absorption of the GaInP window layer

(~1000 nm for sample A and ~50 nm for sample B and C). In any case, only the spectral range above 870 nm is the relevant range for the Ge BC performance in a complete lattice-matched 3JSC. As it can be observed, no major changes appear for wavelengths longer than 1400 nm. On the contrary, a slight difference in the 870-1400 nm range can be appreciated for nucleation B and C with respect to the conservative design (nucleation A).

In line with IQE results, dark J-V curves in Figure *3* evidence a better performance with nucleation routines B and C. In particular, for nucleation B (red dots) a lower recombination current was achieved with respect to nucleation A (black dots). Using a one-diode model with an ideality factor n=1 [30, 31], the fitted recombination currents for sample A, B and C were $J_{01\_A}$= 1.25·10$^{-6}$ A/cm$^2$, $J_{01\_B}$= 0.71·10$^{-6}$ A/cm$^2$ and $J_{01\_C}$= 0.94·10$^{-6}$ A/cm$^2$, respectively. Additionally, Figure *3* also includes an estimation of the open-circuit voltage ($\widetilde{V_{OC}}$), which was also calculated using a one-diode model and the $J_{SC}$s obtained from IQE (Figure 2). From this approach, a $V_{OC}$ gain of ~20 mV is estimated for *Routine B* with respect to the conservative design (*Routine A*).

In light of these results, if we just focus our attention on 1JSCs, we might presumably associate better solar cell performance with both moderate and aggressive nucleation designs developed in this work. But what about the impact of these nucleation routines in complete 3JSCs?

IQEs of the Ge BC subcells of routines A, B and C are presented in Figure *4*. As the upper subcells were identical for the three structures, no changes are expected in their IQEs. Just for reference, the external quantum efficiency (EQE) of both the TC and MC of routines A and B are shown in the inset of Figure *4*. As can be seen in Figure 4, all bottom cell IQEs are very similar and no evident difference stands out for any of the three nucleation routines. In fact, $J_{SC}$ reaches values close to ~18 mA/cm$^2$ in all cases. This value is in agreement with previous works in our group for Ge BCs of complete 3JSCs without anti reflection coating (ARC) layers [5].

IQE results reveal that despite as-grown Ge subcells (1JSCs) show evident differences in Jsc (Figure *2*), the thermal load associated with the growth of the middle and top subcells even out such differences reducing the $J_{SC}$ of the cells to similar final levels, as evidenced by Figure 4. This was also observed during the development of four-junction structures based on Ge bottom cells [32] where similar IQEs of the Ge BC were obtained between 2J and 4J. In conclusion, it does not make sense to optimize the IQE performance of the Ge BC out of a 3JSC structure (i.e; it makes no sense to optimize it in a 1JSC structure that is going to be implemented in a 3JSC).

In a similar way, 3JSCs were analyzed by means of I-V curves. Figure 5 presents the light I-V curves at 500 × of 3JSC devices from each design. The fill factor (FF) and open circuit voltage ($V_{OC}$) shown in Figure 5 confirm that these cells are close to the state-of-the-art of this technology [2]. Noticeable differences in the electrical performance are observed. On the one hand, nucleation A and B present reasonable I-V curves with high fill factors together with a significant gain of 50 mV in $V_{OC}$ for nucleation B. On the other hand, the I-V curve of the 3JSC incorporating nucleation C reveals the presence of a parasitic junction or barrier (blue curve in Figure 5). These results were confirmed in the dark I-V curves in Figure 6 where a lower recombination current is observed when nucleation B is used instead of nucleation A. For nucleation C, a lower recombination current is also observed at low voltages (2-2.5 V) with respect to nucleation A, but this effect is masked at higher voltages (>2.5 V) by the presence of a high series resistance.

Let's consider first the anomalous case resulting from nucleation C. The goal of this routine was to minimize the thermal load of the nucleation process but it turned out with the presence of a high series resistance that ruins the performance of the 3JSC.
This can be explained in terms of the Ge diffusion from the substrate during the epitaxial growth, as already described in several works [33-35]. We have quantified this effect by measuring by SIMS the penetration of Ge atoms into the first two III-V layers for the particular case of nucleation B (Figure 7). This figure shows that Ge diffuses through the 50 nm thick GaInP nucleation layer, penetrating into the Ga(In)As buffer for as much as ~200 nm. For routine B (an even more for A), the ~500 nm thick Ga(In)As buffer layer is more than enough to absorb such diffusion avoiding a significant Ge penetration into the first tunnel junction and middle cell BSF grown immediately above. However, in sample C, we have an extremely thin Ga(In)As buffer layer of only 50 nm (this thickness is represented by the blue rectangle in Figure 7). Consequently, we interpret the effects observed in sample C as the result of Ge solid-phase diffusion reaching the tunnel junction, partially compensating the supposedly high p-type doping of its anode, and thus having a deleterious impact on its properties. This is detected in both dark and light I-V curves, where the presence of a high series resistance or a reverse biased parasitic junction is observed. After this analysis, the bottom line is that when designing very aggressive III-V/Ge nucleation routines, the trade-off between thermal load reduction and mitigation of Ge out-diffusion needs to be carefully tuned.

Moving back to routines A and B, the results presented for these two designs show that the $V_{OC}$ of the 3JSC is notably affected by the different nucleation routines, with a ~50 mV difference between both designs, whereas there is no effect on $J_{SC}$. In this case, we do not have the subcell selectivity provided by the IQE measurement to clarify where the gain in $V_{OC}$ stems from. Even if only ~ 25 mV difference was achieved in 1JSCs, our first hypothesis is that this $V_{OC}$ gain might be mostly originated by the different structures of the bottom cells. In order to discriminate this, we performed EL measurements on average 3JSCs from designs A and B and extracted the individual I-V curves using Uwe Rau's reciprocity theorem [36, 37]. The resulting curves are presented in Figure 8 (black and red circles). A shift towards lower recombination current densities in both the Ge BC and the Ga(In)As MC can be appreciated for sample B. The estimated $V_{OC}$ gain in both subcells was ~ 25 mV each, splitting the $V_{OC}$ gain in almost perfect halves. This relative distance of ~ 25 mV is maintained across a wide range of concentrations (see Figure 8). On the contrary, no significant change is appreciated in the dark I-V curves for the GaInP TC which indicates alike recombination dynamics in both structures. On the side of the Ge BC, we ascribe its improvement in $V_{OC}$ to the reduction of the thermal load during the growth of the nucleation and buffer layers, which possibly improves the quality of the GaInP/Ge interface and of the emitter, as described by Barrigón [5]. Additionally, as reported by Barrigón and co-workers, the thermal load associated with upper subcells degrades the $V_{OC}$ of the Ge BC. Similar relative differences in the Ge BC $V_{OC}$ as those observed in 1JSCs are obtained. Thus, in spite of the aforementioned changes in these two nucleation routines, this second thermal load seems to affect in a similar way both designs. Regarding the Ga(In)As MC, we believe that the improvement comes from a better crystalline quality in this material as a result of the thinner nucleation and buffer layers used. In other words, for nucleation B, the formation of morphological defects such as arrow heads or truncated pyramids [38, 39] might be minimized with thin GaInP nucleation layers of ~50 nm. Additionally, the Ga(In)As buffer layer of ~500 nm smooths out the morphological issues caused by such defects leading to a good template for the growth of the upper III-V layers. Last but not least, the reduction in the growth temperature of the Ga(In)As buffer layer down to 600°C might also help on the formation of such a good template. Future work will be carried out on this point.

The comparison of the so-called conservative, intermediate and aggressive III-V/Ge nucleation routines has made evident that the first stages of the growth have a strong impact on the final 3JSC performance. Despite the differences in $J_{SC}$ are negligible in 3JSCs (for the three particular cases considered in this work) there is a clear impact on $V_{OC}$ or even the presence of high series resistance. Several trade-offs need to be balanced when designing the nucleation routines that form the bottom cell and their final optimization needs to be done on the full 3JSC structure.

**Conclusions**

We have studied the impact of three archetypical III-V/Ge nucleation routines on the growth of high efficiency MJSCs. To this end, single junction Ge solar cells as well as complete triple-junction GaInP/Ga(In)As/Ge solar cells have been grown implementing such three nucleation routines. Differences observed in $J_{SC}$ of single Ge BCs end up vanishing after the growth of complete 3JSCs structures. On the contrary, relative Ge BC $V_{OC}$ differences remain the same either in 1JSC and 3JSC structures. Therefore, when aiming at the optimization of these nucleation routines not only single Ge BCs should be grown but the impact on complete 3JSCs needs to be considered. Additionally, EL and SIMS measurements reveal the important role of these first layers, not only affecting as expected the Ge BC but also having an influence on upper subcells. In particular, the performance of the MC was improved with the so called *moderate* nucleation design described in this study. This was attributed to a better surface crystal quality obtained after this nucleation routine (smoother morphology and lower defect density). On the other hand, if *aggressive* designs with very thin nucleation layers are used, special attention needs to be paid to the Ge diffusion from the substrate that will have a non-negligible impact on upper active layers. In summary, the design of new III-V nucleation routines preserving good morphologies with lower growth times and/or lower temperatures has shown to be of outmost interest in order to maximize the performance of high efficiency MJSCs based on Ge substrates.

**Acknowledgements**

This work has been supported by the Spanish MINECO through the projects TAILLON - TEC2015-66722-R, PCIN-2015-181-C02-02 and TEC2017-83447-P and also by the Comunidad de Madrid through the project MADRID-PV2 (S2018/EMT-4308). Iván García is funded by the Spanish Programa Estatal de Promoción del Talento y su Empleabilidad through a Ramón y Cajal grant (RYC2014-15621). Enrique Barrigón wants to acknowledge funding from the European Union's Horizon 2020 research and innovation programme under the Marie Sklodowska-Curie grant agreement No 656208. This publication reflects only the author's views and the funding agency is not responsible for any use that may be made of the information it contains. The authors want to acknowledge J. Bautista and L. Cifuentes for their technical support and the technical assistance of J. Geisz with EL measurements carried out at NREL.

**Declaration of competing interest**

The authors declare that they have no known competing financial interests or personal relationships that could have appeared to influence the work reported in this paper. As Ignacio Rey-Stolle, a co-author on this paper, is an Editor of SOLMAT, he was blinded to this paper during review, and the paper was independently handled by Dr. Simon P. Philipps.

Table 1: Summary of the three different strategies explored for the growth of both the GaInP nucleation layer and the Ga(In)As buffer.

| Nucleation routine | GaInP nucleation | | | Ga(In)As buffer | | |
|---|---|---|---|---|---|---|
| | Thickness (nm) | T (°C) | Growth time | Thickness (nm) | T (°C) | Growth time |
| A (conservative) | 350 | 675 | 26´45´´ | 1000 | 675 | 20´ |
| B (intermediate) | 50 | 675 | 3´37´´ | 500 | 600 | 10´ |
| C (aggressive) | 50 | 675 | 3´37´´ | 50 | 675 | 1´ |

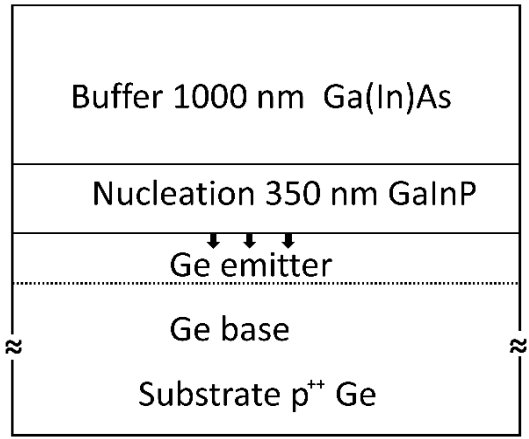

Figure 1: Sketch of the conservative structure (A) of the Ge single junction solar cell grown for this study. Layer thicknesses are not drawn to scale. The emitter is formed by in-diffusion of group-V and group-III elements into the p-type Ge substrate. Likewise, both GaInP nucleation layer and Ga(In)As buffer layer were modified for designs B and C as described in *Table 1*.

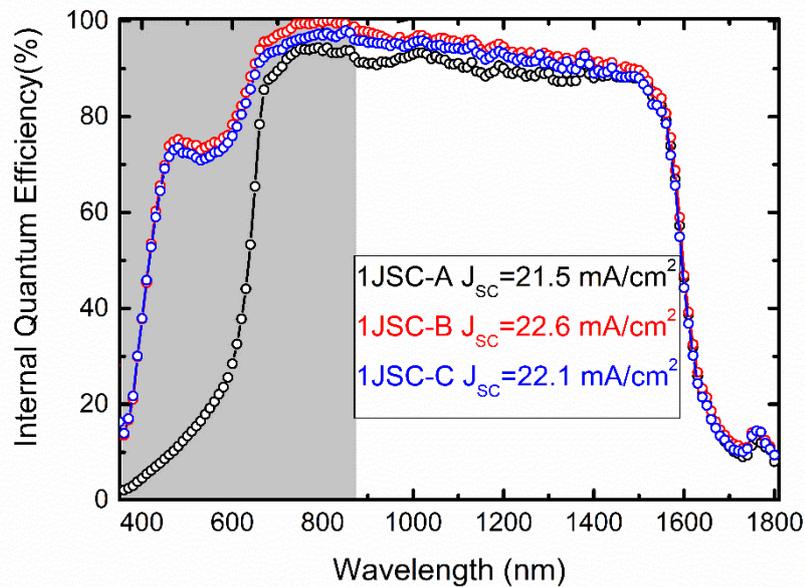

Figure 2: Internal quantum efficiency (IQE) of 1JSCs with nucleation routines A, B and C. The changes observed from 370-685 nm approximately are related to different optical thickness, i.e., absorption of the GaInP window layer (~1000 nm for sample A and ~50 nm for sample B and C). The short circuit current ($J_{SC}$) was calculated from the convolution of the IQE and the AM1.5d standard spectrum along the 870-1800 nm wavelength region which is the relevant range of a Ge BC in a complete 3JSC structure.

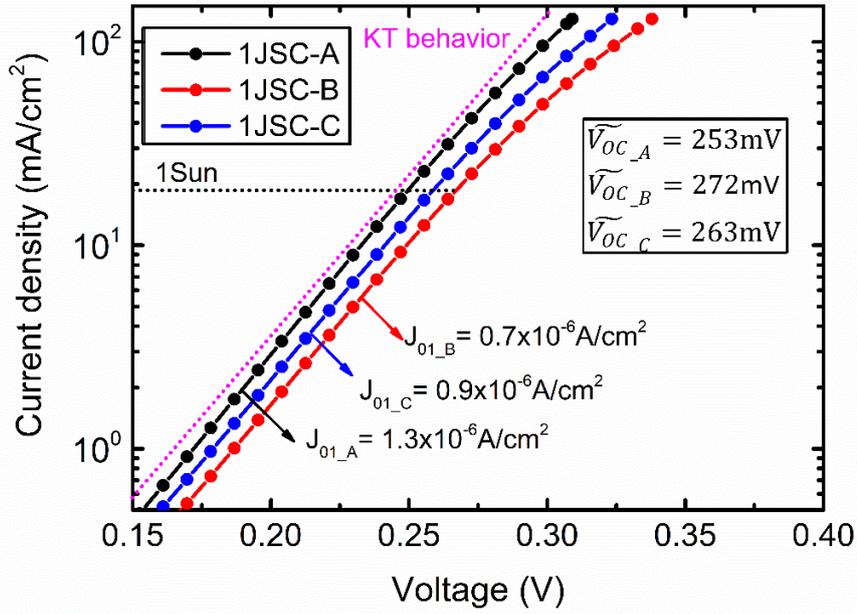

Figure 3: Dark J-V curves of the three different Ge BCs with nucleation profiles A, B and C. The estimated open-circuit voltage ($\widetilde{V_{OC}}$) was calculated at 1sun with the $J_{SC}$ previously obtained from IQE. Recombination current $J_{01}$ was calculated by using the Hovel model approximation to one diode and considering the ideality factor n=1.

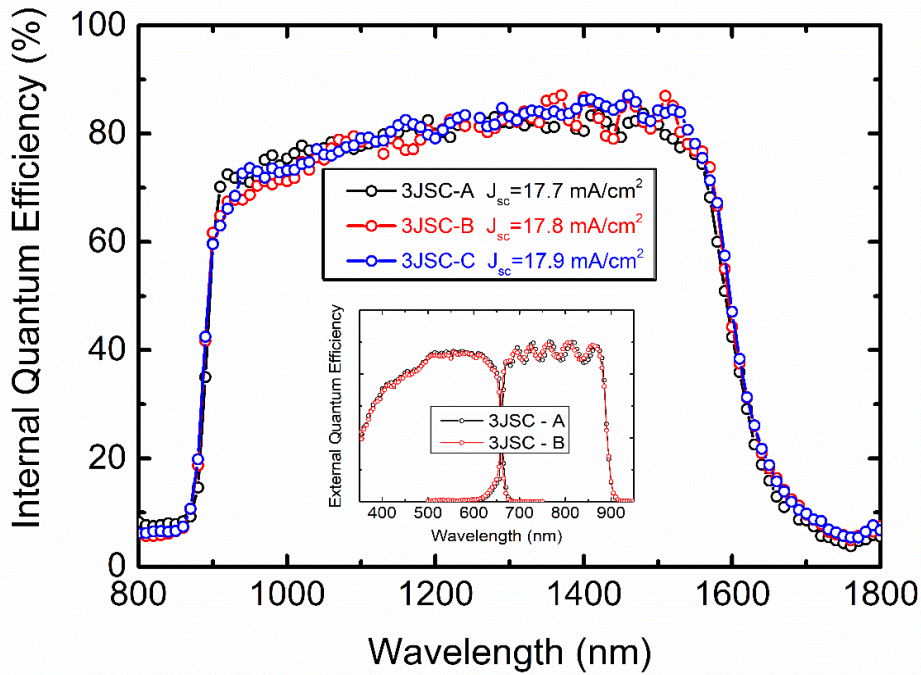

Figure 4: Internal quantum efficiency (IQE) of the Ge subcell in a complete 3JSC structure with nucleation routines A, B and C. No changes are expected in the external quantum efficiency (EQE) of the upper subcells since they were designed in a similar way. As an example, EQE of both TC and MC of 3JSC-A and 3JSC-B are shown in the inset.

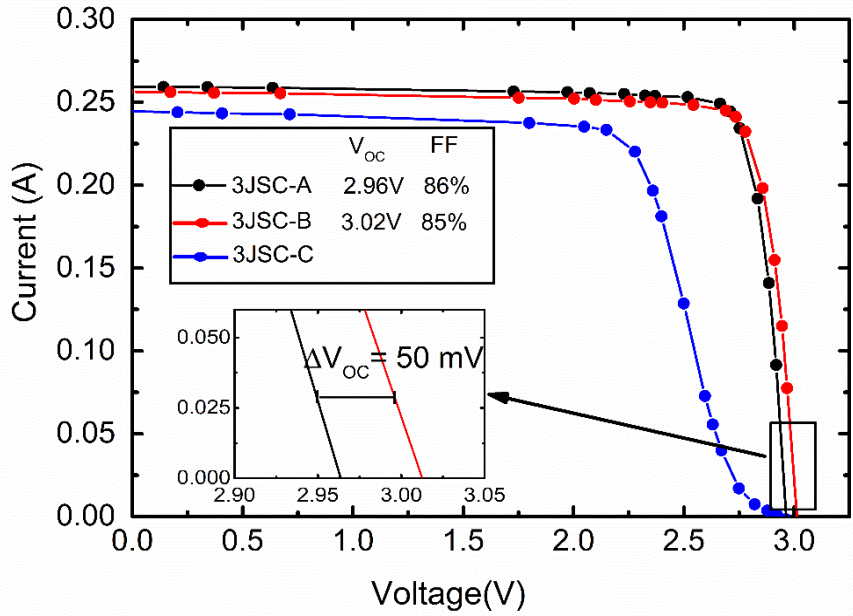

Figure 5: Light I-V curves (~500 suns) of the 3JSC integrating nucleation routine A, nucleation routine B and nucleation routine C. A magnification at high voltages to confirm a $V_{OC}$ gain of 50mV for sample B with respect to the conventional sample A is shown in the inset.

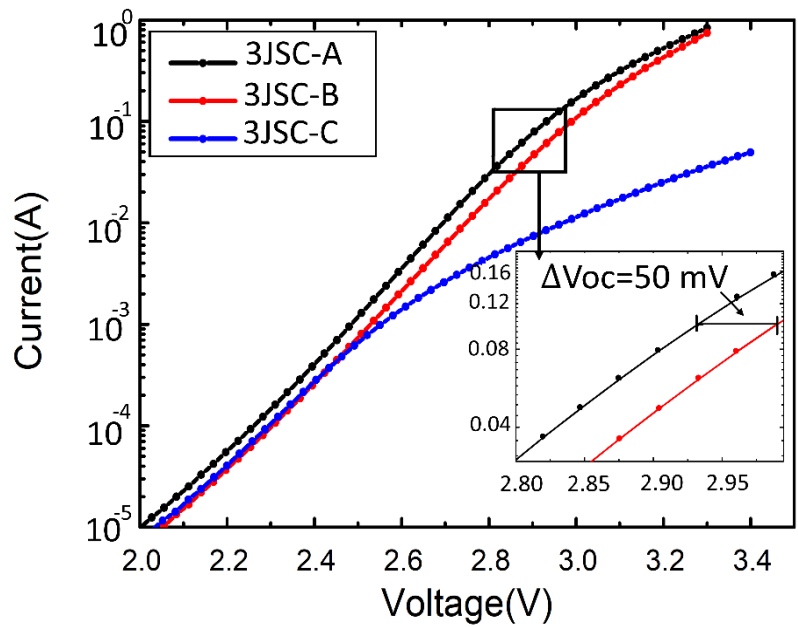

Figure 6: Dark I-V curves of the 3JSC integrating nucleation routines A, B and C. The inset shows a magnified view of the high voltage region, to highlight the $V_{OC}$ gain of 50 mV for sample B as compared to sample A.

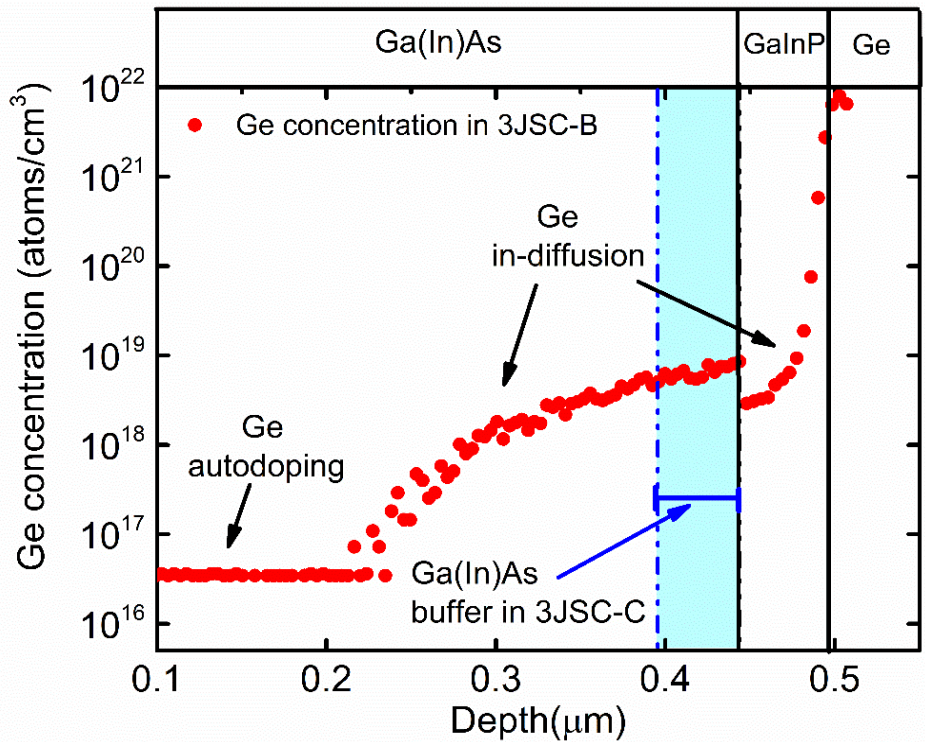

Figure 7: SIMS measurement of the Ge concentration in the first two layers grown on a complete 3JSC- B (~50 nm GaInP and ~500 nm Ga(In)As). Ge in-difussion penetrates the thin GaInP nucleation layer and reaches the Ga(In)As buffer layer. The flat region on the left corresponds to the region where Ge autodoping dominates over Ge in-diffusion. Blue region indicates the very thin Ga(In)As buffer layer designed for 3JSC-C (~50 nm) as compared to 3JSC-B (~500 nm).

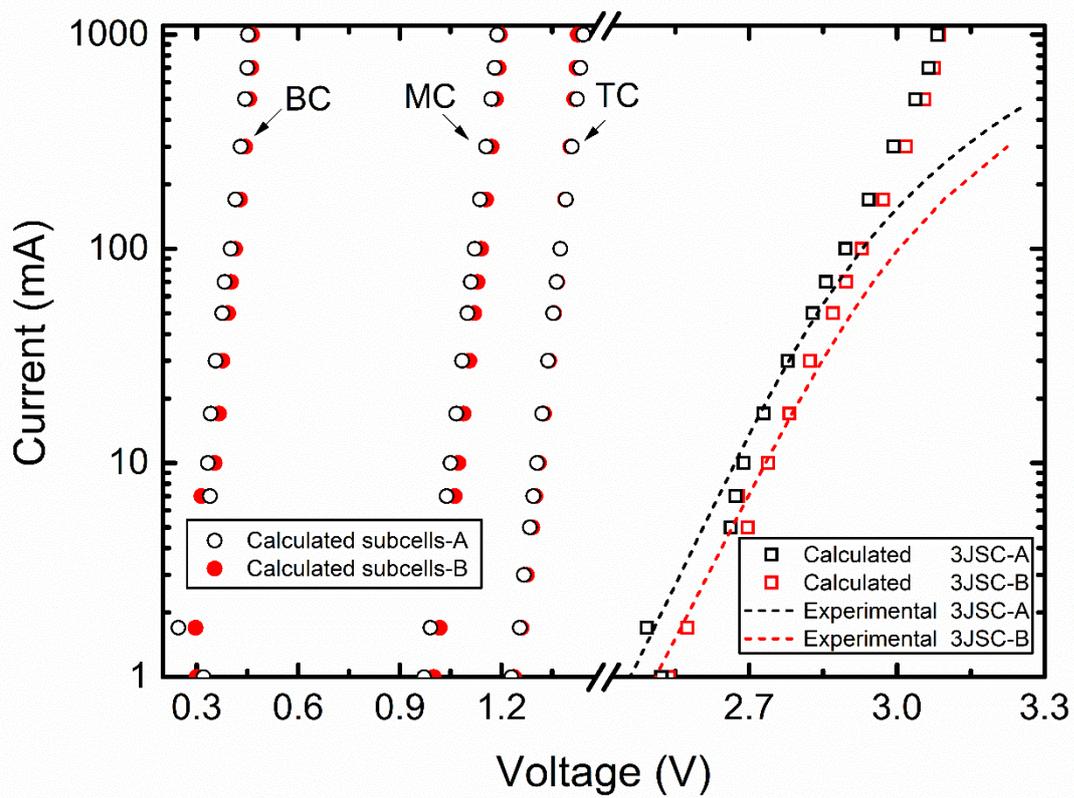

Figure 8: Dark I-V curve of each subcell (bottom cell (BC), middle cell (MC) and top cell (TC)): black open circles correspond to I-V curves from 3JSC-A and red circles to I-V curves from 3JSC-B obtained from EL measurements. The total I-V curves of the GaInP/Ga(In)As/Ge structures directly measured (dash lines) and the total I-V curve reproduced from EL and EQE measurements (squares) are also represented. Due to the low emission detected for the Ge BC during EL measurements, I-V curves results will be considered for currents higher than 10 mA.